\documentclass[10pt]{article}

\usepackage{amssymb}
\usepackage{amsmath}
\usepackage{mathrsfs}
\usepackage{dsfont}
\usepackage{epsfig}

\def \proofn {\noindent {\em Proof. }}
\def \qed{\hfill $\Box$ \vskip 2ex}

\newcommand {\vv} [1]{``#1''}

\newcommand {\ket}[1]{|#1\rangle}
\newcommand  {\bra}[1]{\langle #1|}

\def \Hi{\mathcal{H}}

\def \supp{\textrm{supp}}

\def \C{\mathbb{C}}

\def \R{\mathbb{R}}

\def \Bi{\mathcal{B}}

\def \ddt{\frac{d}{d t}}

\def \Span{\textrm{span}}

\def \diag{\textrm{diag}}

\def \trace{\textrm{trace}}

\def \II{{\mathbb I}}

\def \cal {\mathcal}

\def \beq{\begin{equation}}
\def \eeq{\end{equation}}
\def \beqa{\begin{eqnarray}}
\def \eeqa{\end{eqnarray}}
\def \beqan{\begin{eqnarray*}}
\def \eeqan{\end{eqnarray*}}
\def \bea{\begin{eqnarray}}
\def \eea{\end{eqnarray}}

\newtheorem{definition}{Definition}

\newtheorem{prop}{Proposition}
\newtheorem{thm}{Theorem}
\newtheorem{cor}{Corollary}
\newtheorem{lemma}{Lemma}

\title{Analysis and synthesis of attractive \\
quantum Markovian dynamics}

\author{Francesco Ticozzi\thanks{F. Ticozzi is with the Dipartimento
di Ingegneria dell'Informazione, Universit\`a di Padova, via Gradenigo
6/B, 35131 Padova, Italy ({\tt ticozzi@dei.unipd.it}).} and Lorenza
Viola\thanks{L. Viola is with the Department of Physics and Astronomy,
Dartmouth College, 6127 Wilder Laboratory, Hanover, NH 03755, USA
({\tt Lorenza.Viola@Dartmouth.edu}).}}

\date{\today}

\begin{document}
\maketitle

\begin{abstract} 

We propose a general framework for investigating a large class of
stabilization problems in Markovian quantum systems. Building on the
notions of invariant and attractive quantum subsystem, we characterize
attractive subspaces by exploring the structure of the invariant sets
for the dynamics. Our general analysis results are exploited to assess
the ability of open-loop Hamiltonian and output-feedback control
strategies to synthesize Markovian generators which stabilize a target
subsystem, subspace, or pure-state.  In particular, we provide an
algebraic characterization of the manifold of stabilizable pure states
in arbitrary finite-dimensional Markovian systems, that leads to a
constructive strategy for designing the relevant controllers.
Implications for stabilization of entangled pure states are addressed
by example.
\end{abstract}

\vspace{0.5mm}

\section{Introduction} 

Stabilization problems have a growing significance for a variety of 
quantum control applications, ranging from state preparation of
optical, atomic, and nano-mechanical systems to the generation of
noise-protected realizations of quantum information in realistic
devices \cite{knill-protected,viola-generalnoise}.
Dynamical systems undergoing Markovian evolution
\cite{alicki-lendi,petruccione} are relevant from the standpoint of
typifying irreversible quantum dynamics and present distinctive
control challenges \cite{altafini-markovian}.  In particular,
open-loop quantum-engineering and (approximate) stabilization methods
based on dynamical decoupling cease to be viable in the Markovian
regime \cite{viola-1,viola-engineering}.  It is our goal in this work
to show how a wide class of Markovian stabilization problems can
nevertheless be effectively treated within a general framework,
provided by {\em invariant and attractive quantum subsystems}.

After providing the relevant technical background, we proceed
establishing a first analysis result that fully characterizes the
attractive subspaces for a given generator. This is done by analyzing
the structure induced by the generator in the system's Hilbert space,
and by invoking Krasowskii-LaSalle's invariance principle.  We next
explore the application of the result to stabilization problems for
Markovian Hamiltonian and output-feedback control. Our approach leads
to a complete characterization of the stabilizable pure states,
subspaces, and subsystems as well as to constructive design strategies
for the control parameters. Some partial results in this sense have
been presented in \cite{ticozzi-QDS}, and in the conference paper
\cite{ticozzi-QDSPHYSCON}. We also refer to the journal article
\cite{ticozzi-QDS} for a more detailed discussion of the connection
between invariant, attractive and noiseless subsystems, along with a
thorough analysis of model robustness issues which shall not be our
focus here.

\section{Preliminaries and background}

\subsection{Quantum Markov processes}

Consider a separable Hilbert space $\Hi$ over the complex field $\C$.
Let ${\mathfrak B}(\Hi)$ represent the set of linear bounded operators
on $\Hi$, with ${\mathfrak H}(\Hi)$ denoting the real subspace of
Hermitian operators, and ${\mathbb I}$, ${\mathbb O}$ being the
identity and the zero operator, respectively.  Throughout our
analysis, we consider a {\em finite-dimensional} quantum system
$\mathcal{Q}$: following the standard quantum statistical mechanics
formalism \cite{sakurai}, we associate to $\mathcal{Q}$ a complex,
finite-dimensional $\Hi$.  Our (possibly uncertain) knowledge of the
state of $\mathcal{Q}$ is condensed in a {\em density operator} $\rho$
on $\Hi$, with $\rho\geq 0$ and $\trace(\rho)=1$.  Density operators
form a convex set ${\mathfrak D}(\Hi)\subset {\mathfrak H}(\Hi)$, with
one-dimensional projectors corresponding to extreme points ({\em pure
states}, $\rho_{|\psi\rangle}=|\psi\rangle\langle \psi|$). Observable
quantities are represented by Hermitian operators in ${\mathfrak
H}(\Hi),$ and expectation values are computed by using the trace
functional: $\mathbb{E}_\rho(X)=\trace(\rho X).$ If ${\cal Q}$ is the
composite system obtained from two distinguishable quantum systems
${\cal Q}_1,\,{\cal Q}_2$, the corresponding mathematical description
is carried out in the tensor product space,
$\Hi_{12}=\Hi_1\otimes\Hi_2$, observables and density operators being
associated with Hermitian and positive-semidefinite, normalized
operators on $\Hi_{12}$, respectively.  The {\em partial trace} over
$\Hi_2$ is the unique linear operator $\trace_{2}(\cdot): {\mathfrak
B}(\Hi_{12})\rightarrow {\mathfrak B}(\Hi_{1})$, ensuring that for
every $X_1\in{\mathfrak B}(\Hi_{1}),X_2\in{\mathfrak B}(\Hi_{2})$,
$\trace_{2}(X_1\otimes X_2)=X_1 \trace(X_2)$.  Partial trace is used
to compute marginal states and partial expectations on multipartite
systems.

In the presence of either intended or unwanted couplings (such as with
a measurement apparatus, or with a surrounding quantum environment),
the evolution of a subsystem of interest is no longer unitary and
reversible, and the general formalism of {\em open quantum systems} is
required \cite{davies,alicki-lendi,petruccione}.  A wide class of open
quantum systems obeys Markovian dynamics
\cite{alicki-lendi,lindblad,qds,petruccione}. Let ${\cal I}$ denote
the physical quantum system of interest, with associated Hilbert space
$\Hi_I$, $\textrm{dim}(\Hi_I)=d$.  Assume that we have no access or
control over the state of the system's environment, and that the
dynamics in ${\mathfrak D}(\Hi_I)$ is continuous in time and described
at each instant $t\geq 0$ by a Trace-Preserving Completely Positive
(TPCP) linear map ${\cal T}_t(\cdot)$ \cite{kraus}.  If a forward
composition law is also assumed, we obtain a quantum Markov process,
or Quantum Dynamical Semigroup (QDS):
\begin{definition}[QDS]\label{QDS} 
A {\em quantum dynamical semigroup} is a one-parameter family of TPCP
maps $\{{\cal T}_t(\cdot),\,t\geq0\}$ that satisfies:
\begin{itemize}
\item  ${\cal T}_0={\mathbb I}$, 
\item ${\cal T}_t\circ{\cal T}_s={\cal T}_{t+s},\; \forall t,s >0,$  
\item $\textrm{{\em trace}}({\cal T}_t(\rho) X)$
is a continuous function of $t$, $\forall\rho\in{\mathfrak D}(\Hi_I)$,
$\forall X\in \Bi(\Hi_I)$.
\end{itemize}
\end{definition}
\noindent 
Due to the trace- and positivity-preserving assumptions, a QDS is a
semigroup of contractions.  As proven in 
\cite{lindblad,gorini-k-s}, the Hille-Yoshida generator for the 
semigroup exists and can be cast in the following
canonical form:
\begin{eqnarray}
\dot{\rho}(t)&=&\mathcal{L}(\rho(t))=
-\frac{i}{\hbar}[H,\rho(t)]+\sum_{k=1}^{p}\gamma_k {\cal
D}(L_k,\rho(t))\,\label{eq:lindblad}\\
&=&-\frac{i}{\hbar}[H,\rho(t)]+\sum_{k=1}^{p}
{\gamma_k}\Big(\hspace*{-1mm}L_k\rho(t)L_k^\dag-\frac{1}{2}\{L_k^\dag
L_k,\,\rho(t)\}\hspace*{-1mm}\Big), \nonumber
\end{eqnarray}
with $\{\gamma_k\}$ denoting the spectrum of $A$. The {\em effective
Hamiltonian} $H$ and the {\em noise operators} $L_k$ (also known as
``Lindblad operators'') completely specify the dynamics, including the
effect of the Markovian environment.  In general, $H$ is equal to the
Hamiltonian for the isolated, free evolution of the system, $H_0$,
plus a correction, $H_L$, induced by the coupling to the environment
(aka ``Lamb shift'').  The non-Hamiltonian terms ${\cal
D}(L_k,\rho(t))$ in \eqref{eq:lindblad} account for the non-unitary
character of the dynamics, specified by noise operators $\{L_k\}$.

In principle, the exact form of the generator of a QDS may be
rigorously derived from a Hamiltonian model for the joint
system-environment dynamics under appropriate limiting conditions (the
so-called ``singular coupling limit'' or the ``weak coupling limit,''
respectively~\cite{alicki-lendi,petruccione}).  In most physical
situations, however, an analytical derivation is unfeasible, since the
full microscopic Hamiltonian describing the system-environment
interaction is unavailable.  A Markovian generator of the form
\eqref{eq:lindblad} is then postulated on a phenomenological basis. In
practice, it is often the case that knowledge of the noise effect may
be assumed, allowing to specify the Markovian generator by directly
assigning a set of noise operators $L_k$ (not necessarily orthogonal
or complete) in \eqref{eq:lindblad}, and the corresponding noise
strengths $\gamma_k$.  Each of the noise operators $L_k$ may be
associated to a distinct {\em noise channel} ${\cal D}(L_k,\rho(t))$,
by which information irreversibly leaks from the system to the
environment.

\subsection{Quantum subsystems: Invariance and attractivity }

Quantum subsystems are the basic building block for describing
composite systems in quantum mechanics~\cite{sakurai}, and provide a
general framework for scalable quantum information engineering in
physical systems. In fact, the so-called {\em subsystem principle}
\cite{viola-generalnoise,knill-protected,viola-ips} states that any
``faithful'' representation of information in a quantum system
requires to specify a subsystem 
desired information. Many of the control tasks considered in this
paper are motivated by the need for strategies to create and maintain
quantum information in open quantum systems.  A definition of quantum
subsystem suitable to our scopes is the following:

\begin{definition}[Quantum subsystem] \label{def:subsys}
A {\em quantum} {\em subsystem} $\mathcal{S}$ of a system
$\mathcal{I}$ defined on $\Hi_I$ is a quantum system whose state space
is a tensor factor $\Hi_S$ of a subspace $\Hi_{SF}$ of $\Hi_I$,
\beq\Hi_I =\Hi_{SF}\oplus \Hi_R= (\Hi_{S}\otimes\Hi_F)\oplus \Hi_R,
\label{eq:subs}\eeq
\noindent
for some co-factor ${\cal H}_F$ and remainder space ${\cal H}_R$.  The
set of linear operators on ${\cal S}$, ${\cal B}({\cal H}_S)$, has the
same statistical properties and is isomorphic to the (associative) 
subalgebra of ${\cal B}(\Hi_I)$ of operators of the form $X_I=X_{S}\otimes
 \mathbb{I}_F \oplus {\mathbb O}_R$.
\end{definition}

Let $n=\dim(\Hi_S),$ $f=\dim(\Hi_F),$ $r=\dim(\Hi_R)$, and let
$\{\ket{\phi_j^{S}}\}_{j=1}^n,\,\{\ket{\phi_k^{F}}\}_{k=1}^f,$ 
$\{\ket{\phi_l^{R}}\}_{l=1}^r$ be orthonormal bases for
$\Hi_{S},\,\Hi_F,\,\Hi_R,$ respectively.  The decomposition
\eqref{eq:subs} is then naturally associated with the following basis
for $\Hi_I$:
$$\{\ket{\varphi_m}\}=\{\ket{\phi_j^{S}}
\otimes\ket{\phi_k^{F}}\}_{j,k=1}^{n,f}\cup\{\ket{\phi_l^{R}}\}_{l=1}^r.$$
\noindent
This induces a block structure for matrices acting on $\Hi_{I}$:
\beq\label{eq:blocks} X=\left(
\begin{array}{c|c}
  X_{SF} & X_P     \\\hline
  X_Q  &  X_R   
\end{array}
\right), \eeq 
\noindent
where, in general, $X_{SF}\neq X_{S}\otimes X_F$. We denote by
$\Pi_{SF}$ the projector onto $\Hi_{SF}$, that is,
$\Pi_{SF}=\left(\begin{array}{c|c} \mathbb{I}_{SF} &
0\end{array}\right)$.

In this paper, we study Markov dynamics of a quantum system ${\cal I}$
with a given decomposition of the associated Hilbert space of the form
\eqref{eq:subs}, with respect to the quantum subsystem $\cal S $
associated to $ \Hi_S$.  By describing the dynamics in the
Schr\"odinger picture, {\em i.e. }, with evolving states and
time-invariant observables, the first step is to specify whether the
system ${\cal I}$ has been properly initialized in a state which
faithfully represents a state of the subsystem ${\cal S}$, and what is
the structure of such states.

\begin{definition}[State initialization]\label{initialization} 
The system $\mathcal{I}$ with state $\rho\in{\mathfrak D}(\Hi_{I})$ is
{\em 
initialized in $\Hi_{S}$ with state $\rho_{S}
\in{\mathfrak D}(\Hi_{S})$} if the blocks of $\rho$ satisfy:

\begin{tabular}{l}
\hspace*{-2mm}{\em (i)} $\rho_{SF}=\rho_{S}\otimes\rho_F$ for some
$\rho_{F}\in{\mathfrak D}(\Hi_{F});$ \\
\hspace*{-2mm}{\em (ii)} $\rho_P=0,\rho_R=0.$
\end{tabular}

\noindent
We denote by ${\mathfrak I}_S(\Hi_I)$ the set of states that satisfy
{\em (i)-(ii)} for some $\rho_S$.
\end{definition}

Condition (ii) guarantees that $\bar{\rho}_S=\trace_F(\Pi_{SF}\rho
\Pi_{SF}^\dag)$ is a valid (normalized) state of ${\cal S}$, while
condition (i) ensures that measurements or dynamics affecting the
factor $\Hi_F$ have no effect on the state in $\Hi_S$.

We now proceed to characterize in which sense, and under which
conditions, a quantum subsystem may be defined as invariant. Recall
that a set ${\cal W}$ is said to be {\em invariant} for a dynamical
system if the trajectories that start in ${\cal W}$ remain in ${\cal
W}$ for all $t\geq 0$.\footnote{For clarity, let us also recall other
standard dynamical systems notions relevant in our context. Given
$\dot\rho={\mathcal L}(\rho)$ and a suitable norm for the state
manifold, we call {\em invariant, stationary, or equilibrium state}
any $\rho$ such that $\dot\rho=0.$ An equilibrium state $\rho$ is said
to be {\em stable} if for every $\epsilon\geq 0$ there exists $\delta$
such that if $\|\rho_0-\rho\|\leq\delta$, then any trajectory starting
from $\rho_0$ does not leave the ball of radius $\epsilon$ centered in
$\rho$. A state $\rho$ is said to be (globally) {\em attractive} if
the trajectories from any initial condition converge to it.}  In view
of Definition \ref{initialization}, the natural definition considering
dynamics in the state space may be phrased as follows:

\begin{definition}[{Invariance}] \label{invariance} 
Let ${\cal I}$ evolve under a family of TPCP maps $\{{\cal
T}_t,\,t\geq 0\}$.  ${\cal S}$ is an {\em invariant subsystem} if
${\mathfrak I}_S(\Hi_I)$ is an invariant subset of ${\mathfrak
D}(\Hi_I)$.
\end{definition}

In explicit form, as given in \cite{ticozzi-QDS}, this means that
$\forall\,\rho_{S}\in{\mathfrak D}(\Hi_{S}),\,\rho_{F}\in{\mathfrak
D}(\Hi_{F})$, the state of ${\cal I}$ obeys
\begin{equation}\label{eq:invariance}
{\cal T}_t\left(
\begin{array}{c|c}
  \rho_{S}\otimes\rho_F & 0    \\\hline
  0  &  0  
\end{array}
\right)=\left(
\begin{array}{c|c}
  {\cal T}^S_t(\rho_{S})\otimes{\cal T}^F_t(\rho_F) & 0    \\\hline
  0  &  0  
\end{array}
\right),\, t\geq0,
\end{equation}
where, for every $t\geq 0,$ ${\cal T}^S_t(\cdot)$ and ${\cal
T}^F_t(\cdot)$ are TPCP maps on $\Hi_S$ and $\Hi_F$, respectively,
{not} depending on the initial state.  For Markovian evolution of
$\mathcal{I}$, $\{{\cal T}^S_t\}$ and $\{{\cal T}^F_t\}$ are required
to be QDSs on their respective domain.

We next recall a characterization of dynamical models able to ensure
invariance for a fixed subsystem, based on appropriately constraining
the block-structure of the matrix representation of the operators
specifying the Markovian generator.  We refer to \cite{ticozzi-QDS}
for the proofs.

\begin{lemma}[{Markovian invariance}]\label{invariancecondition} 
Assume that $\Hi_I=(\Hi_{S}\otimes\Hi_F)\oplus \Hi_R$, and let
$H,\,\{L_k\}$ be the Hamiltonian and the error generators of a
Markovian QDS as in \eqref{eq:lindblad}. Then $\Hi_{S}$ supports an
invariant subsystem iff $\forall\, k$:
\begin{eqnarray} 
&&L_{k}=\left(
\begin{array}{c|c}
  L_{S,k}\otimes L_{F,k} & L_{P,k} \nonumber \\\hline 0 & L_{R,k}
\end{array}
\right), \\ && iH_P-\frac{1}{2}\sum_k(L_{S,k}^\dag\otimes
L_{F,k}^\dag)L_{P,k}=0,\label{eq:nonrobust}\\ && H_{SF}=H_S\otimes
\mathbb{I}_F +\mathbb{I}_S\otimes H_{F},\nonumber 
\end{eqnarray} 
\noindent 
where for each $k$ either $L_{S,k}=\mathbb{I}_S$ or
$L_{F,k}=\mathbb{I}_F$ (or both). 
\end{lemma}

One may require ${\cal S}$ to have dynamics independent from evolution
affecting ${\cal H}_F$ and ${\cal H}_R$ also in the case where ${\cal
I}$ is not {initialized} in the sense of Definition
\ref{initialization}. If neither conditions (i)-(ii) are satisfied,
one may still define an unnormalized reduced state for the subsystem:
$$\tilde{\rho}_S=\trace_F(\Pi_{SF}\rho\Pi_{SF}^\dag),\quad
\trace(\tilde\rho_S)\leq 1.$$ 
\noindent
This allows for entangled mixed states to be supported on $\Hi_{SF},$
as well as for blocks $\rho_P,\,\rho_R$ to differ from zero.
Similar to the case of so-called \vv{initialization-free} subsystems
considered in \cite{lidar-initializationDFS,ticozzi-QDS}, an
additional constraint on the Lindblad operators is required to ensure independent reduced dynamics in this case.  That is, with respect to the matrix block-decomposition above, it must be $L_{P,k}=0$ for every $k$. Such
a constraint decouples the evolution of the $SF$-block of the state
from the rest, rendering both ${\mathfrak I}_S(\Hi_I)$ and ${\mathfrak
I}_R(\Hi_I)$ {\em separately} invariant.

This imposes tighter conditions on the noise operators, which may be
hard to ensure in reality and, from a control perspective, leave less
room for Hamiltonian compensation as examined in Section
\ref{compensation}.  In order to address situations where such extra
constraints cannot be met, as well as a question which is interesting
on its own, we explore conditions for a subsystem to be {\em
attractive}:

\begin{definition}[{Attractive subsystem}] \label{attraction} 
Assume that $\Hi_I=(\Hi_{S}\otimes\Hi_F) \oplus \Hi_R$. Then $\Hi_{S}$
supports an {\em attractive subsystem} with respect to a family
$\{{\cal T}_t\}_{t\geq 0}$ of TPCP maps if $\forall\rho\in{\mathfrak
D}(\Hi_{I})$ the following condition is asymptotically obeyed:
\begin{equation}\label{eq:attraction}
\lim_{t\rightarrow \infty}\left({\cal T}_t(\rho)-\left(
\begin{array}{c|c}
  \bar\rho_{S}(t)\otimes\bar\rho_F(t) & 0    \\\hline
  0  &  0  
\end{array}
\right)\right)=0,\end{equation} 
\noindent
where $$\bar\rho_{S}(t)={\text{\rm trace}}_{F}[\Pi_{SF}{\cal T}_t(\rho)
\Pi_{SF}^\dag],$$ 
$$\bar\rho_{F}(t)={\text{\rm trace}}_{S}[\Pi_{SF}{\cal T}_t(\rho)
\Pi_{SF}^\dag].$$
\end{definition}

This implies that every trajectory in ${\frak D}(\Hi_I)$ converges to
${\frak I}_S(\Hi_I).$ Thus, an attractive subsystem may be thought of
as a subsystem that ``self-initializes'' in the long-time limit, by
reabsorbing initialization errors. Although such a desirable behavior
only emerges asymptotically, for QDSs one can see that convergence is
exponential, as long as the relevant eigenvalues of ${\cal L}$ have
strictly negative real part.

We conclude this section by recalling two partial results on
attractive subsystems which we established in \cite{ticozzi-QDS}.  The
first is a negative result, which shows, in particular, how the
possibility of ``initialization-free'' and attractive behavior are 
mutually exclusive.

\begin{prop}\label{hermitianL} Assume that
$\Hi_I=(\Hi_{S}\otimes\Hi_F)\oplus \Hi_R,$ $\Hi_R\neq 0$, and let
$H,\,\{L_k\}$ be the Hamiltonian and the error generators as in
\eqref{eq:lindblad}, respectively.  Let $\Hi_{S}$ support an invariant
subsystem. If $L_{P,k}=L_{Q,k}^\dag=0$ for every $k$, then $\Hi_{S}$
is not attractive.\end{prop}

Note that he conditions of the above Proposition are obeyed, in
particular, if $L_k=L_P^\dag$, $\forall k$.  As a consequence,
attractivity is never possible for the class of {\em unital}
(${\mathbb I}$-preserving) Markovian QDSs with purely self-adjoint
$L_k$'s.  Still, even if the condition $L_{P,k}=L_{Q,k}^\dag=0$
condition holds, attractive subsystems may exist in the pure-factor
case, where $\Hi_R=0$. Sufficient conditions are provided by the
following:

\begin{prop}\label{attractionfactor} Assume that
$\Hi_I=\Hi_{S}\otimes\Hi_F$ ($\Hi_R= 0$), and let $\Hi_{S}$ be
invariant under a QDS with generator of the form 
$${\cal L}= {\cal L}_{S}\otimes {\mathbb I}_F + {\mathbb I}_S \otimes{\cal
  L}_F .$$ 
\noindent
If ${\cal L}_F(\cdot)$ has a unique attractive state $\hat\rho_F$,
then $\Hi_S$ is attractive. \end{prop}

Interesting linear-algebraic conditions for determining whether a
generator ${\cal L}_F(\cdot)$ has a unique attractive state (though
not necessarily pure) are presented in
\cite{spohn-equilibrium,spohn-approach}.

\section{Characterizing attractive Markovian dynamics}

We begin by presenting new {\em necessary and sufficient} conditions
for attractivity of a subspace, which will provide the basis for the
synthesis results in the next sections. Notice that if $\Hi_{SF}$
supports an attractive subsystem, the entire set of states with
support on $\Hi_{SF},$ ${\mathfrak I}_{\cal SF}(\Hi_{I}),$ is
attractive. Once this is verified, the dynamics confined to the
invariant subspace (that supports a pure subsystem) may be studied
with the aid of the results recalled in the previous section.  The
following Lemma will be used in the proof of the main result, but is
also interesting on its own. We denote with $\supp(X)$ the support of
$X\in{\mathfrak B}(\Hi),$ {\em i.e.}, the orthogonal complement of its
kernel.

\begin{lemma}\label{support}
Let ${\cal W}$ be an invariant subset of ${\frak D}(\Hi_I)$ for the
QDS dynamics generated by $\dot\rho={\cal L} (\rho),$ and define:
$$\Hi_{\cal W}=\mbox{\em supp} ({\cal W})=\bigcup_{\rho\in{\cal
W}}\mbox{\em supp} (\rho).$$ Then ${\frak I}_{\cal W}(\Hi_I)$ is
invariant.
\end{lemma}
\proofn Let $\hat{\cal W}$ be the convex hull of ${\cal W}.$ Thus,
every element $\hat \rho$ of $\hat{\cal W}$ may be expressed as
$\hat\rho = \sum_k p_k \rho_k,$ where $p_k\geq 0,$ $\sum_k p_k=1$, and
$\rho_1,\ldots,\rho_k\in{\cal W}.$ By using linearity of the dynamics,
$${\cal T}_t(\hat\rho) = \sum_k p_k {\cal T}_t(\rho_k)=\sum_k p_k
\rho'_k, \;\;\; \forall t \geq0.$$ with
$\rho_1',\ldots,\rho'_k\in{\cal W}.$ Hence $\hat{\cal W}$ is
invariant.  Furthermore, from the definition of $\hat{\cal W}$, there
exist a $\bar\rho\in \hat{\cal W}$ such that
$\supp(\bar\rho)=\supp(\hat{\cal W})=\Hi_{\cal W}.$ Consider
$\Hi_I=\Hi_{\cal W}\oplus \Hi^{\perp}_{\cal W}, $ and the
corresponding matrix partitioning:
$$X=\left(
\begin{array}{c|c}
X_{\cal W} & X_P\\ \hline
X_Q & X_R\\
\end{array}
\right).$$ With respect to this partition, the block $\bar\rho_{\cal
W}$ of $\bar \rho$ is full-rank, while $\bar\rho_{P,Q,R}$ are
zero-blocks.  The trajectory $\{{\cal T}_t(\bar\rho),t\geq 0\}$ is
contained in $\hat{\cal W}$ only if:
$$\ddt\bar\rho=\left(
\begin{array}{c|c}
{\cal L}_{\cal W}(\bar\rho_{\cal W}) & 0\\ \hline
0 & 0\\
\end{array}
\right),$$ so that, upon computing explicitly the generator blocks, we
must impose:
\begin{eqnarray*}
\hspace*{1.5cm}\left\{\begin{array}{l}
\bar\rho_{\cal W}\left(iH_P-\frac{1}{2}\sum_kL_{{\cal W},k}^\dag L_{P,k}\right)=0,\\
-\frac{1}{2}\sum_k\{L^\dag_{Q,k}L_{Q,k},\,\bar\rho_{\cal W}\}=0.
\end{array}\right.
\end{eqnarray*}
Since $\bar\rho_{\cal W}$ is full-rank and positive, it must be: 
\begin{eqnarray*}
\hspace*{1.5cm}\left\{\begin{array}{l}
iH_P-\frac{1}{2}\sum_k L_{{\cal W},k}^\dag L_{P,k}=0,\\
L_{Q,k}=0,\quad\forall k.
\end{array}\right.
\end{eqnarray*}
Comparing with the conditions given in Corollary 
\ref{invariancecondition}, we infer that 
${\frak I}_{\cal W} (\Hi_I)$ is invariant, hence we conclude.
\qed

We are now in a position to prove our main result:
  
\begin{thm}[{Subspace attractivity}]
\label{attractivity} Let $\Hi_I=\Hi_S\oplus\Hi_R,$ and assume that 
$\Hi_S$ is an invariant subspace for the QDS dynamics generated by
\eqref{eq:lindblad}.  Define: \beq\label{Rprime} \Hi_{R'}=
\bigcap_{k=1}^{p} \ker(L_{P,k}), \eeq with the matrix blocks $L_{P,k}$
representing linear operators from $\Hi_R$ to $\Hi_S.$ Then $\Hi_{S}$
is an {\em attractive} subspace iff $\Hi_{R'}$ does not
support any invariant subsystem.
\end{thm}

\proofn Clearly, if $\Hi_{R'}$ supports an invariant set ${\cal W}_R$,
then $\Hi_{S}$ cannot be attractive, since for every $
\bar\rho\in{\cal W}_R,$ the dynamics is confined to ${\cal W}_R.$ To
prove the other implication, we shall prove that if $\Hi_{R'}$ does
not support an invariant {\em set}, then $\Hi_S$ is attractive.
Consider the non-negative, linear functional $V(\rho)=\trace(\Pi_R
\rho).$ It is zero iff $\rho_R=0$, {\em i.e.}, for perfectly
initialized states. By LaSalle's invariance principle (see {\em e.g.}
\cite{khalil-nonlinear}), every trajectory will converge to the
largest invariant subset ${\cal W}$ contained in the set:
$${\cal Z}=\{\rho\in {\frak D}(\Hi_I)|\dot V(\rho)=0\}.$$ Explicit
calculation of the blocks of the generator (see also
\cite{ticozzi-QDS}) yields:
$$\dot V( \rho)=\trace(\Pi_R {\cal
L}(\rho))=-\trace\Big(\sum_kL_{P,k}^\dag L_{P,k}\rho_R\Big).$$ 
\noindent
By the cyclic property of the trace, the last term is equivalent to
the trace of $\sum_k L_{P,k}\rho_R L_{P,k}^\dag,$ which is a sum of
positive operators, and thus can be zero iff each term is zero. Being
the $L_{P,k}$'s fixed, this can hold iff $\rho_R$ has support
contained in $\Hi_{R'}$, defined as above.  Thus, the support of
${\cal Z}$ is $\Hi_S\oplus\Hi_{R'}.$ Call $\Hi_{\cal W}$ the support
of the maximal invariant set ${\cal W}$ in ${\cal Z}$: By Lemma
\ref{support}, ${\frak I}_{\cal W}(\Hi_I)$ is invariant. But ${\cal
W}$ is defined as the maximal invariant set in ${\cal Z}$, so it must
be ${\cal W}={\frak I}_{\cal W}(\Hi_I).$ Recalling that by hypothesis
${\frak I}_S( \Hi_I )$ is itself an invariant subset contained in
${\cal Z}$, it must be $\Hi_{\cal W}= \Hi_S\oplus\Hi_{R''},$ with
$\Hi_{R''}\subset \Hi_{R'}.$ We next prove that $\Hi_{R'}$ supports an
invariant set iff $\Hi_{\cal W}\neq\Hi_{S},$ {\em i.e.}  $\Hi_{R'}\neq
0$\footnote{To the scope of this proof, the ``if'' implication would
suffice, but since the converse arises naturally, we prove both.}.
Consider a $\hat\rho\in{\cal W}$ such that $\hat\rho$ has non-trivial
support on $\Hi_{S}^\perp.$ If no such state exists, ${\cal W}$ has
support only on $\Hi_S,$ so clearly $\Hi_{R'}=0.$ If such a state
exists, let
$$\hat\rho'=
\frac{\Pi_{R'}\hat\rho\Pi_{R'}}{\trace(\Pi_{R'}\hat \rho)},$$ where
$\Pi_{R'}$ is the orthogonal projector on $\Hi_{R'}.$ Since
$\hat\rho'$ has support only on $\Hi_{R'}\subset \Hi_{\cal W},$ its
trajectory $\{\hat\rho'(t)={\cal T}_t(\hat\rho'), t\geq 0\}$ is
confined to ${\cal W}.$ On the other hand, ${\cal W}\subset {\cal Z},$
hence it must be $\dot V(\hat\rho(t))=0$ for all $t\geq0.$ By
observing that $V(\hat\rho')=1$ and that $V(\rho),\,\dot V(\rho)$ are
continuous, we can conclude that the trajectory $\{\hat \rho'(t)\}$
must have support only on some $\Hi_{R''}\subset\Hi_{R'},$ endowing
$\Hi_{R'}$ with an invariant set, and by the Lemma above, the
invariant subsystem associated to ${\frak I}_{R''}(\Hi_I)$.  We
conclude by observing that if $\Hi_{R'}$ does not support an invariant
set, then $\Hi_{\cal W}=\Hi_{S},$ hence $\Hi_S$ is attractive.  \qed

In spite of its non-constructive nature, the power of this
characterization will be apparent in the proofs of the results
concerning {\em active stabilization} of states and subspaces by
Hamiltonian control in the next section.  We observe here that Lemma
\ref{support} lends itself to the following useful specialization:
\begin{prop}\label{supportstate}
If $\bar\rho$ is an invariant state 
for the QDS
dynamics generated by $\dot\rho={\cal L}(\rho),$ and $\Hi_B=\supp
(\bar\rho),$ then ${\frak I}_B(\Hi_I)$ is invariant. Conversely, if
$\bar\Hi$ supports an invariant subset, it contains at least an
invariant state.
\end{prop}
\proofn The first implication follows from Lemma \ref{support}
above. If $\bar\Hi$ supports an invariant subset, then by the same
Lemma it supports an invariant subsystem, and the density operators
with support on $\bar\Hi$ form a convex, compact set that evolves
accordingly to a (reduced) QDS. Hence, it must admit at least an
invariant state \cite{alicki-lendi}.  \qed

This result provides us with an explicit criterion for verifying
whether $\Hi_{R'}$ contains an invariant subset: It will suffice to
check if $\Hi_{R'}$ supports an invariant state. Invariant (or
``fixed'') states may be found by analyzing the structure of
$\ker({\cal L}(\cdot))$. An efficient algorithm for generic TPCP maps
has been recently presented in \cite{viola-ips}.

\section{Engineering attractive Markovian dynamics}

In this section, we illustrate the relevance of the theoretical
framework developed thus far to a wide class of Markovian
stabilization problem associated with the task of making a desired
(fixed) quantum subsystem invariant or attractive. Interestingly,
these problems may be regarded as instances of {\em Markovian
reservoir engineering}, which has long been investigated on a
phenomenological basis by the physics community in the context of both
decoherence mitigation and the quantum-classical transition, see {\em
e.g.}  \cite{poyatos,davidovich,dematos}.

In the special yet relevant case of sinthesizing attractive dynamics
with respect to an intended {\em pure} state, our results fully
characterize the manifold of pure states that may be ``prepared''
given a reference dissipative dynamics using either open-loop
Hamiltonian or feedback control resources.  As discussed in
\cite{ticozzi-QDS}, provided a sufficient level of accuracy in tuning
the relevant control parameters may be ensured, the ``direct''
Markovian feedback considered here has the important advantage of
substantially relaxing implementation constraints in comparison with
``Bayesian'' feedback techniques requiring real-time state estimation
update \cite{doherty-linear,vanhandel-feedback}.

\subsection{Open-loop Hamiltonian control}
\label{compensation}

We begin by exploring what can be achieved by considering only
open-loop Hamiltonian control, specifically, the application of {\em
time-independent} Hamiltonians to the dynamical generator. This allows
us to consider generators involving, in general, multiple $L_k,$ and
yields interesting characterizations of the possibilities offered by
this class of controls for stabilization problems, complementing
previous work from a controllability perspective
\cite{altafini-open,schirmer-controllability}. The results established
below will also be of key importance in the proofs of the theorems on
closed-loop stabilization.  Lastly, a separate presentation will serve
to clarify the different scopes and limitations of the two class of
control strategies.  As a direct consequence of the Markovian
invariance theorem, we have the following:

\begin{cor}[Open-loop invariant subspaces]\label{openloopinvariant} 
Let $\Hi_I=\Hi_S\oplus\Hi_R.$ Then ${\frak I}_S(\Hi_I)$ can be made
invariant by open-loop Hamiltonian control 
iff $L_{Q,k}=0$ for every $k$.
\end{cor}
\proofn By specializing Corollary \ref{invariancecondition}, $\Hi_{S}$
supports an invariant subsystem iff:
\begin{eqnarray} 
&&L_{Q,k}=0,\quad \forall\, k \label{cQ}\\ 
&& iH_P-\frac{1}{2}\sum_k L_{S,k}^\dag L_{P,k}=0.\label{subspaceinterplay}
\end{eqnarray} 
The only condition that is affected by a change of Hamiltonian is
\eqref{subspaceinterplay}, which however can always be satisfied by an
appropriate choice of control Hamiltonian. This leaves us with
condition \eqref{cQ} alone.  \qed

The above result makes it possible to enforce invariant subspaces for
the controlled dynamics by solely using Hamiltonian resources, {\em
without directly modifying the non-unitary part}. The ability of
open-loop Hamiltonian control to induce stronger attractivity
properties is characterized in the following:
\begin{thm}[Open-loop attractive subspaces]
\label{openloopattractivity} Let 
$\Hi_I=\Hi_S\oplus\Hi_R$ and assume that $\Hi_S$ supports an invariant
subsystem. Then ${\frak I}_S(\Hi_I)$ can be made attractive by
open-loop Hamiltonian control iff ${\frak I}_{R}(\Hi_{I})$ is not
invariant.
\end{thm}
\proofn If $\Hi_R$ supports an invariant subsystem, then by Corollary
\ref{invariancecondition} it must be $L_{P,k}=0$ for every $k$.  Since
$\Hi_S$ invariant, this implies $H_P=0.$ Any Hamiltonian control
perturbation that preserves invariance on $\Hi_S$ must satisfy this
condition, hence preserve invariance on $\Hi_R$ too, thus $\Hi_S$
cannot be rendered attractive.  If the whole $\Hi_R$ does not support
an invariant subsystem, we can devise an iterative procedure that
builds up a control Hamiltonian $H_c$ such that $\Hi_S$ becomes
attractive.  Theorem \ref{attractivity} states that if there is no
invariant subsystem supported in $\Hi_{R'}$ (defined in
\eqref{Rprime}), then $\Hi_S$ is attractive.  If there is an invariant
subsystem with support $\Hi_T\subset \Hi_{R'},$ let us consider the
following Hilbert space decomposition: $$\Hi_I= \Hi_S\oplus
\Hi_{T}\oplus \Hi_{Z}.$$ After imposing the invariance conditions on
$\Hi_S$ and $ \Hi_T,$ the associated block-decomposition of the
Lindblad operators and Hamiltonian turns out to be of the form:
$$L_{k}=\left(
\begin{array}{c|c|c}
  L_{S,k} & 0 & L_{P',k} \nonumber \\ \hline 
  0 & L_{T,k} & L_{P'',k} \\ \hline
  0 & 0 & L_{Z, k}
\end{array}
\right), $$
$$H=\left(
\begin{array}{c|c|c}
  H_{S} & 0 & H_{P'} \nonumber \\ \hline 
  0 & H_{T} & H_{P''} \\ \hline
  H^\dag_{P'} & H^\dag_{P''} & H_{Z}
\end{array}
\right), $$
subject to the conditions:
\begin{eqnarray*}
\hspace*{2cm}iH_{P'}-\frac{1}{2}\sum_k L_{S,k}^\dag L_{P',k}=0,\\
\hspace*{2cm}iH_{P''}-\frac{1}{2}\sum_k L_{T,k}^\dag L_{P'',k}=0.
\end{eqnarray*} One sees that
the most general Hamiltonian perturbation that preserves the
invariance of $\Hi_S$ has the form:
$$H_c=\left(
\begin{array}{c|c|c}
  H_{1} & 0 & 0 \nonumber \\ \hline 
  0 & H_{2} & H_{M} \\ \hline
 0 & H^\dag_{M} & H_{3}
\end{array}
\right). $$ Consider a control Hamiltonian $H_c$ such that the block
$H_M$ has full column-rank, while $H_1,H_3$ are arbitrary and $H_2$ is
still to be determined. If $\dim(\Hi_T)\leq \frac{1}{2}\dim(\Hi_R),$
then $i\rho_T H_M\neq 0$ for every $ \rho_T,$ hence $\Hi_T$ cannot
support any invariant subsystem, since conditions in Corollary
\ref{invariancecondition} cannot be satisfied for any subspace of
$\Hi_T$.  Conversely, if $\dim(\Hi_T)> \frac{1}{2}\dim(\Hi_R),$
choosing an $H_M$ as above, by dimension comparison $H_M$ must have a
non-trivial left kernel ${\cal K}$, ${\cal K}H_M=0,$ and thus there
exists a $\Hi_{T'}\subset{\cal K}$ that supports an invariant ${\frak
I}_{T'}(\Hi_I),$ whose dimension is strictly lesser than dimension of
$\dim(\Hi_T).$ We can iterate the reasoning with a new, refined
decomposition $\Hi_I= \Hi_S\oplus \Hi_{T'}\oplus \Hi_{Z'},$ with $
\Hi_{Z'}= \Hi_{Z}\oplus(\Hi_{T}\ominus \Hi_{T'})$. With this
decomposition, the generator matrices exhibits the same block
structure as above, with $\dim(\Hi_{T'})<\dim(\Hi_{T}).$ Thus, we can
exploit the freedom of choice on the block $H_2$ to further reduce the
dimension of the invariant set.  At each iteration, the procedure
either stops rendering $\Hi_S$ attractive, if $\dim(\Hi_T)\leq
\frac{1} {2}\dim(\Hi_R),$ or decrease the dimension of the invariant
set by at least $1.$ The procedure thus ends in at most
$\dim(\Hi_{R'})$ steps.  \qed


Remarkably, the proof of the above theorem, combined with a strategy
to find invariant subspaces, provides a constructive procedure to
build a constant Hamiltonian that makes the desired invariant
subspace attractive whenever the Theorem's hypothesis are satisfied.

\subsection{Markovian feedback control}
\label{markoviancontrol}

The potential of Hamiltonian compensation for controlling Markovian
evolutions is clearly limited by the impossibility to directly modify
the noise action. To our scopes, open-loop control is then mostly
devoted to connect subspaces in $\Hi$ that are already invariant, and
to adjust the generator parameters so that the interplay between
Hamiltonian and dissipative contributions (as in
Eq. \eqref{eq:nonrobust}) can stabilize the desired subspace or
subsystem.

A way to overcome these limitations is offered by closed-loop control
strategies.  Measurement-based feedback control requires the ability
to both effectively monitor the environment, and condition the target
evolution upon the measurement record. Feedback strategies have been
considered since the beginning of the quantum control field
\cite{belavkin-towards}, and successfully employed in a wide variety
of settings (see {\em e.g.}
\cite{wiseman-milburn,wiseman-feedback,doherty-linear,mabuchi-science,ticozzi-feedbackDD}).

We focus on a measurement scheme which mimics optical homo-dyne
detection for field-quadrature measurements, whereby the target system
({\em e.g.} an atomic cloud trapped in an optical cavity) is
indirectly monitored via measurements of the outgoing laser field
quadrature \cite{wiseman-milburn,thomsen}. The conditional dynamics of
the state is stochastic, driven by the fluctuation one observes in the
measurement.  Considering a suitable infinitesimal feedback operator
determined by a {\em feedback Hamiltonian} $F$, and taking the
expectation with respect to the noise trajectories, this leads to the
Wiseman-Milburn Markovian {\em Feedback Master equation} (FME)
\cite{wiseman-milburn,wiseman-feedback}: \beq\label{eq:MME}
\dot{\rho}_t=-i\hbar\Big[H+ \frac{1}{2}(FM+M^\dag F),\,\rho_t \Big]+{\cal
D}(M - iF,\rho_t).\eeq
\noindent

The feedback state-stabilization problem for Markovian dynamics has
been extensively studied for a single two-level system ({\em qubit})
~\cite{wang-wiseman,wiseman-bayesian}. The standard approach is to to
design a Markovian feedback loop by assigning {both} the measurement
and feedback operators $M,F,$ and to treat the measurement strength
and the feedback gain as the relevant control parameters
accordingly. Throughout the following section, we will assume to have
more freedom, by considering, for a fixed measurement operator $M$,
{\em both} $F$ and $H$ as tunable control Hamiltonians

\begin{definition} [CHC] A controlled FME of the form \eqref{eq:MME} 
supports {\em complete Hamiltonian control} (CHC) if {\em (i)}
arbitrary feedback Hamiltonians $F\in {\mathfrak H}(\Hi_I)$ may be
enacted; {\em (ii)} arbitrary {\em constant} control perturbations
$H_c\in {\mathfrak H}(\Hi_I)$ may be added to the free Hamiltonian
$H$.
\end{definition}

This leads to both new insights and constructive control protocols for
systems where the noise operator is a generalized angular
momentum-type observable, for generic finite-dimensional systems.
Physically, the CHC assumption must be carefully scrutinized on a case
by case basis, since constraints on the form of the Hamiltonian with
respect to the Lindblad operator may emerge, notably in the
abovementioned weak-coupling limit derivations of Markovian
models~\cite{alicki-lendi}.

We now address the general subspace-stabilization problem for
controlled Markovian dynamics described by FMEs.  A characterization
of the subspaces supporting stabilizable subsystems is provided by the
following:


\begin{thm} [Feedback attractive subspaces]\label{subspaceattractivity}
Let $\Hi_I=\Hi_S\oplus \Hi_R$, with $\Pi_S$ being the orthogonal projection
on $\Hi_S.$ Assume CHC capabilities.  Then, for any measurement
operator $M$, there exist a feedback Hamiltonian $F$ and a Hamiltonian
compensation $H_c$ that make the subsystem supported by $\Hi_S$
attractive for the FME \eqref{eq:MME} iff
\beq\label{stabilize}[\Pi_S,(M+M^\dag)]
\neq0.\end{equation}
\end{thm} 
\proofn Write $M=M^H+i M^A$, with both $M^H$ and $M^A$ being
Hermitian, thus $L=M^H+i(M^A-F)$.  Condition \eqref{stabilize} holds
iff $M^H$ is not block-diagonal when partitioned according to the
chosen decomposition.  If $M^H$ is block-diagonal, then, by Corollary
\ref{invariancecondition}, enforcing invariance of the subsystem
supported by $\Hi_S$ requires that $L_Q=0$.  But then it must also be
$L_P=0,$ so that $\Hi_R$ supports an invariant subsystem.  Since the
choice of $L_S$ and $L_R$ does not affect invariance, by Theorem
\ref{openloopattractivity} it follows that $\Hi_S$ cannot be made
attractive by Hamiltonian control.  On the other hand, if $M^H$ is not
block-diagonal, we can always find $F$ in such a way that $L$ is upper
diagonal, $L_P\neq 0,$ by choosing $F_P=iM^H_P+M^A_P$.  With $L$ as
the new noise operator, we now have to devise a control Hamiltonian
$H_c$ with a block $H_{c,P}$ that makes
$\Hi_S$ invariant (this is always possible by Corollary
\ref{openloopinvariant}, since $L_Q$=0), and a block $H_{c,R}$
constructed following the procedure in the proof of Theorem
\ref{openloopattractivity}. \qed

The following specialization to pure states, {\em i.e.}
one-dimensional subspaces, is immediate:

\begin{cor} \label{purestateattractivity}
Assume CHC. For any measurement operator $M$, there exist a feedback
Hamiltonian $F$ and a Hamiltonian compensation $H_c$ able to stabilize
an arbitrary desired pure state $\rho_d$ for the FME \eqref{eq:MME}
iff 
\beq\label{stabilize2}[\rho_d,(M+M^\dag)]
\neq0.\end{equation}
\end{cor} 

The proof of Theorem \ref{subspaceattractivity} provides a
constructive algorithm for designing the feedback and correction
Hamiltonians needed for the stabilization task.  In particular, our
analysis recovers the qubit stabilization results of
\cite{wang-wiseman} recalled before.  For example, the states that are
not stabilizable within the control assumptions of \cite{wang-wiseman}
are the ones commuting with the Hermitian part of $M=\sigma_+,$ that
is, $M^H=\sigma_x.$ In the $xz$-plane of the Bloch's representation,
the latter correspond precisely to the equatorial points.

As a corollary of Theorem \ref{subspaceattractivity} and Proposition
\ref{attractionfactor}, we present sufficient and necessary conditions
for engineering a generic attractive quantum subsystem (with a
non-trivial co-factor). We start with a Lemma, which is a
straightforward specialization of Proposition 5 in \cite{ticozzi-QDS}:

\begin{lemma}\label{2states} Assume that 
$\Hi_I=\Hi_{S}\otimes\Hi_F$, ($\Hi_R= 0$), 
and a QDS of the form
${\cal L}= {\mathbb I}_{S}\otimes{\cal L}_F.$
If ${\cal L}_F(\cdot)$ admits at least two invariant states, then
$\Hi_S$ is not attractive. \end{lemma}

\begin{thm}[Feedback attractive subsystems] \label{subsystemattractivity}
Let $\Hi_I=\Hi_{SF}\oplus\Hi_R=\Hi_S\otimes\Hi_F\oplus \Hi_R,$ with
$\dim(\Hi_S),\dim(\Hi_F)\geq 2,$ and assume CHC capabilities.  Then
for any $M$, with Hermitian part $M^H$, there exist a feedback
Hamiltonian $F$ and a Hamiltonian compensation $H_c$ that make the
subsystem ${\cal S}$ attractive for the FME \eqref{eq:MME} iff
the following conditions hold: 
\begin{enumerate}\item[{\em i)}] \vspace*{-2mm}
\beq\label{stabilize3}  \vspace*{-2mm}[\Pi_{SF},M^H]
\neq0, \end{equation} 

\item[{\em ii)}]  
\beq\label{Msf} \vspace*{-2mm}\Pi_{SF}M^H\Pi_{SF}^\dag=\left\{
\begin{array}{l}\II_S\otimes C_F,\;\text{\rm or} \\ C_S\otimes \II_F,
\end{array}\right.\eeq

\item[{\em iii)}] 
\beq\label{Msf1}\vspace*{-2mm} \Pi_{SF}M^H\Pi_{SF}^\dag\neq \lambda
\II_{SF},\; \forall \lambda\in \C. \eeq
\end{enumerate}
\end{thm} 

\proofn By Theorem \ref{subspaceattractivity}, condition
\eqref{stabilize3} is necessary and sufficient to render $\Hi_{SF}$
attractive, which is a necessary condition for attractivity of
$\Hi_S.$ In fact, if this is not the case, by Theorem
\ref{attractivity} there would exist an invariant subsystem whose
support is contained $\Hi_R.$ To ensure invariance of ${\frak
I}_S(\Hi_I),$ by Corollary \ref{invariancecondition}, the block
$L_{SF}$ of $L=M-iF$ has to satisfy $L_{SF}=L_S\otimes L_F,$ with
$L_{S}=\mathbb{I}_S$ or $L_{F}=\mathbb{I}_F$ (or both). Thus, both the
Hermitian and anti-Hermitian parts of $L_{SF}$ must have the same
structure. The Hermitian part of $L$ is equal the Hermitian part of
$M$, whereby it follows that
$\eqref{Msf}$ is necessary for invariance of ${\frak
I}_S(\Hi_I)$. Assume $C_F\neq\II_F$ (the other case may be treated in
a similar way, by interchanging the roles of $\Hi_F$ and $\Hi_S$ in
what follows). If $\eqref{Msf1}$ is not satisfied, then $L_{SF}$ must
be unitarily similar to a diagonal matrix for any choice of $F$ that
ensures invariance of ${\frak I}_S(\Hi_I)$. Hence, the dynamics
restricted to $\Hi_F$ admits at least two different stationary states
($\dim(\Hi_F)\geq 2$ by hypothesis). By Lemma \ref{2states}, we
conclude that ${\frak I}_S(\Hi_I)$ cannot be attractive.  Conversely,
if i) holds, following the proof of Theorem
\ref{subspaceattractivity}, we can devise a Hamiltonian correction
$H_c$ and a feedback Hamiltonian $F$ for which $\Hi_{SF}$ is
attractive. Since the $SF$-block is irrelevant to this stage, $H_c$
and $F$ may be further chosen to render a pure state of $\Hi_F$
attractive for the reduced dynamics. Assume ii) and iii), with $C_F$
different from a scalar matrix (again, to treat the other case,
$C_S\neq\lambda \II_S,$ it suffices to switch the appropriate
subscripts in what follows). Thus, there exists a one-dimensional
projector $\rho_{1}$ such that $[\rho_{1},C_F] \neq 0.$ By Corollary
\ref{purestateattractivity}, we can find $F_F$ and $H_F$ that render
it attractive. By choosing an Hamiltonian control so that
$H_{SF}=\II_S\otimes H_F,$ and $F_{SF}=\II_S\otimes F_F,$ the stated
conditions are also sufficient for the existence of
attractivity-ensuring controls.  \qed

\section{Applications} 

The following examples will serve to exemplify the application of our
stabilization results to prototypical finite-dimensional control
systems, which are also of direct relevance to quantum information
devices.  Different scenarios may arise depending on whether the
target system is (or is regarded as) indecomposable, or explicit
reference to a decomposition into subsystems is made.

\subsection{Single systems}

{\bf Example 1:} Consider a single qubit on $\Hi\simeq \C^2$, with
uncontrolled dynamics specified by $H=n_0 \mathbb I_2 + n_x\sigma_x +
n_y\sigma_y + n_z\sigma_z$, with $n_0,n_x,n_y, n_z \in \R$ and
$M=\frac{\hbar}{2}\sigma_x.$ Assume we wish to stabilize $\rho_d=\diag
(1,0).$ Since $[\rho_d, \sigma_x]\neq 0,$ this is possible.  Following
the procedure in the above proof, consider
$F=-\frac{\hbar}{2}\sigma_y,$ so that
$$L=\hbar \frac{\sigma_x + i \sigma_y}{2} =\hbar \sigma_+=\hbar \left(%
\begin{array}{cc}
  0 & 1 \\
  0 & 0 \\
\end{array}
\right),$$ and $H_c=-n_x\sigma_x -n_y \sigma_y.$ Substituting in the
FME \eqref{eq:MME}, one obtains the desired result, as it can also be
directly verified by using Proposition 7 in \cite{ticozzi-QDS}.

Assume, more generally, that it is possible to continuously monitor an
arbitrary single-spin observable, $\vec{\sigma}\cdot\vec{n}$.  Since
the choice of the reference frame for the spin axis is conventional,
by suitably adjusting the relative orientation of the measurement
apparatus and the sample, it is then in principle possible to prepare
and stabilize any desired pure state with a similar control strategy.

{\bf Example 2:} Consider a three-level system (a {\em qutrit}), whose
Hilbert space $\Hi \simeq \C^2$ carries a spin-1 representation of
spin angular momentum observables $J_\alpha$, $\alpha=x,y,z$.  Without
loss of generality, we may choose a basis in $\Hi$ such that the
desired pure state to be stabilized is $\rho_d=\textrm{diag}(1,0,0)$,
and by CHC we may also ensure that $H=0$.  In analogy with Example 1,
a natural strategy is to continuously monitor a non-diagonal spin
observable, for instance:
$$ J_x = \frac{\hbar}{\sqrt{2}} 
\begin{pmatrix} 0&1&0\\ 1&0&1\\ 0&1&0 \end{pmatrix}. $$ 
\noindent 
Since $[\rho_d, J_x]\neq 0,$ the state is stabilizable. Choosing 
the feedback Hamiltonian as
$$F=-J_y = -\frac{\hbar}{\sqrt{2}}
\begin{pmatrix} 0&-i&0\\ i&0&-i\\ 0&i&0 \end{pmatrix}, $$ 
yields 
$$ L =J_x +i J_y = \frac{\hbar}{\sqrt{2} } \begin{pmatrix} 0&1&0\\
0&0&1\\ 0&0&0 \end{pmatrix}. $$ 
Unlike the qubit case, $H'=\frac{1}{2}(FM+M^\dagger F) \neq 0$,
thus a Hamiltonian compensation $H_c$ is needed to ensure that
$i(H'+H_c)_P -\frac{1}{2} L^\dagger_S L_P=0$.  With these choices, it
is easy to see that $\Hi_{R'}$ does not support any invariant
subsystem, hence $\rho_d$ is attractive.

Provided that a similar structure of the observables is ensured, the
previous examples naturally extend to generic $d$-level systems, as
formally established in \cite{ticozzi-QDS} by using Lyapunov
techniques.  

\subsection{Bipartite systems}

If a multipartite structure is specified on $\Hi$, it is both
conceptually and practically important to understand whether
stabilization of physically relevant class of states (including
non-classical {\em entangled states}) is achievable with control
resources which respect appropriate operational constraints, such as
locality.  We focus here on the simplest setting offered by {\em
bipartite qubit systems}, with emphasis on Markovian-feedback
preparation of entangled states, which has also been recently analyzed
within a quantum filtering approach in \cite{yamamoto-twospin}.

{\bf Example 3:} Consider a two-qubit system defined on a Hilbert
space $\Hi \simeq\C^2\otimes\C^2$, with a preferred basis ${\cal C}=\{
\ket{ab}=\ket{a}\otimes\ket{b}\,|\,a,b=0,1\}$ ({\em e.g.}, ${\cal C}$
defines the {\em computational basis} in quantum information
applications).  The control task is to engineer a QDS generator that
stabilizes the maximally entangled \vv{cat state}:
$$\rho_d=\frac{1}{2}(\ket{00}+\ket{11})(\bra{00}+\bra{11}).$$ In order
to employ the synthesis techniques developed above, we consider a
change of basis such that in the new representation
$\rho_d=\diag(1,0,0,0).$ A particularly natural choice is to consider
the so-called {\em Bell basis}:
\bea 
&&\hspace*{1cm}{\cal B}=\left\{\frac{1}{\sqrt{2}}(\ket{00}+\ket{11}),
\frac{1}{\sqrt{2}}(\ket{00}-\ket{11}),\right.\nonumber\\
&&\hspace*{2cm} \left.\frac{1}{\sqrt{2}}(\ket{01}+
\ket{10}),\frac{1}{\sqrt{2}}(\ket{01}-\ket{10})\right\}.
\nonumber\eea
\noindent
Let $U$ be the unitary matrix realizing the change of basis. In the
Bell basis, which we use to build our controller, we consider a
Hilbert space decomposition $\Hi=\Hi_S\oplus \Hi_R,$ where
$\Hi_S=\Span\{\ \frac{1}{\sqrt{2}}(\ket{00}+\ket{11})\},$ and
$\Hi_R=\Hi_S^\perp,$ and the associated matrix block decomposition.

Let us consider $M=\sigma_z\otimes\II$ in the canonical basis. It is
easy to verify that $[M,\rho_d]\neq 0.$ In the Bell basis, $M^B=UM
U^\dag=\II\otimes \sigma_x,$ and $M_P=(0,1,0,0).$ If, in this basis,
we are able to implement the feedback Hamiltonian
$F^B=\II\otimes\sigma_y$ (where now the tensor product should simply
be meant as a matrix operation), we render $\rho_d$ invariant, yet
obtaining $L^B=M^B-iF^B$ with $L^B_{P}\neq 0.$ Direct computation
yields $F=U^\dag(\II \otimes \sigma_y) U= \sigma_y\otimes \sigma_x,$
back in the computational basis.  With this choice, using the
definitions in the proof of Theorem \ref{openloopattractivity}, we
have:
$$\Hi_{R'} =\Span
\left\{\frac{1}{\sqrt{2}}(\ket{01}+\ket{10}),
\frac{1}{\sqrt{2}}(\ket{01}-\ket{10})\right\},$$
\noindent
and $\Hi_{R'}$ is itself invariant. Hence, we need to produce a
control Hamiltonian $H_c$ to \vv{destabilize} $\Hi_{R'}.$ By
inspection, we find that $\Hi_{R'}$ contains a proper subspace
$\Hi_T=\Span\{ \frac{1}{\sqrt{2}}(\ket{01}+\ket{10})\}$ that supports
an invariant and attractive state for the dynamics reduced to
$\Hi_{R'}$. To \vv{connect} this state to the attractive domain of
$\rho_d,$ we need a non-trivial Hamiltonian coupling between
$\frac{1}{\sqrt{2}}(\ket{00}-\ket{11})$ and
$\frac{1}{\sqrt{2}}(\ket{01}+\ket{10})$. This may be obtained by a
control Hamiltonian $H_c=\sigma_y\otimes \II + \II \otimes \sigma_y$
in the standard basis -- which completes the specification of the
control strategy that renders $\rho_d$ the unique attractive state for
the dynamics.  Notice that both the measurement and Hamiltonian
compensation can be implemented locally, which may be advantageous in
practice.

This example suggests how our results, obtained under CHC assumptions,
may be interesting to explore the compatibility with existing control
constraints. A further illustration comes from the following example.

{\bf Example 4:} Consider again the above two-qubit system, but now
imagine that we can only implement \vv{non-selective} measurement and
control Hamiltonians, {\em i.e.}, $M,F,H$ must commute with the
operation that swaps the qubit states. It is then natural to restrict
attention to the dynamics in the three-dimensional subspace generated
by the {\em triplet} states, which correspond to eigenvalue $\hbar^2
J(J+1)$, $J=1$, of the total spin angular momentum $J_{\alpha} =
\frac{\hbar}{2}( \sigma_{\alpha} \otimes \II +\II \otimes
\sigma_{\alpha} )$, $\alpha=x,y,z$ \cite{sakurai}:
$$\Hi_{J=1}=
\Span\left\{\ket{00},\frac{1}{\sqrt{2}}(\ket{01}+\ket{10}),\ket{11}\right\}.$$
Notice that $\Hi_{J=1}$ corresponds to the fixed subspace with respect
to the swap operation.

Our goal is to engineer a FME such that the maximally entangled state
$\rho_d=\frac{1}{{2}}(\ket{01}+\ket{10})(\bra{01}+\bra{10})$ is
attractive for the dynamics restricted to $\Hi_{J=1}$.  Consider a
collective measurement of spin along the $x$-axis, described by
$J_x$. Upon reordering the triplet vectors so that in the new (primed)
basis the $z$-projection ranges over $0, 1, -1$ and
$\rho_d=\diag(1,0,0),$ we have:
$$J'_x=\frac{\hbar}{\sqrt{2}}\begin{pmatrix} 0&1&1\\
1&0&0\\ 1&0&0
\end{pmatrix}. $$
\noindent  
Thus, in this basis we are looking for a feedback Hamiltonian of the
form:
$$F'=\frac{\hbar}{\sqrt{2}}
\begin{pmatrix} \;\;0&i&i\\ -i&0&0\\ -i&0&0 \end{pmatrix}, $$
that is, $F'$ corresponds to the non-selective operator $F=\frac{1}{\hbar} (J_z
J_y+J_yJ_z).$ Hence, by choosing $M'=J_x$ and $F'$ as above we get:
$$L'= M'-iF'=\sqrt{2}\hbar \begin{pmatrix} 0&1&1\\ 0&0&0\\ 0&0&0
\end{pmatrix}. $$ 
\noindent
This, with a choice of $H'_c=J_z$ suffices to make $\rho_d$
attractive.  In fact, considering the Hilbert space decomposition
$\Hi_{J=1}=\Hi_S\oplus \Hi_R,$ where $\Hi_S=\Span\{\frac{1}{\sqrt{2}}
(\ket{01}+\ket{10})\},$ and $\Hi_R=\Hi_S^\perp,$ we find that the
largest invariant subset in $\Hi_R$ has support in $\Hi_{R'}
=\Span\{\frac{1}{\sqrt{2}}(\ket{00}-\ket{11})\}.$ By observing that
$\frac{1}{\sqrt{2}}(\ket{00}-\ket{11})$ is not an eigenstate of $J_z$,
we infer that the chosen control parameters make $\rho_d$ attractive
in $\Hi_{J=1}$.


\section{Conclusion}
\label{sec:conclusions}

We have revisited the fundamental concepts of invariance and
attractivity for quantum Markovian subsystems from a system-theoretic
viewpoint. Building on the characterization of invariant subsystems
and some partial results presented in \cite{ticozzi-QDS}, a
linear-algebraic approach and Lyapunov's stability theory methods have
provided us with an explicit characterization of attractive subspaces,
along with an explicit attractivity test. In the special case of a
single pure state, our results directly characterize the semigroup
generators which support {\em state-preparation via dissipative
Markovian dynamics}.

In the second part of the work, the conditions identified for
subsystem invariance and attractivity have been exploited for
designing Hamiltonian and output-feedback Markovian control strategies
which actively achieve the intended quantum stabilization. In addition
to a complete characterization of subspaces and subsystems that can be
rendered attractive, our results include constructive recipes for
synthesizing the required control parameters, which have been
illustrated in simple yet paradigmatic examples.  While our present
analysis assumes perfect detection efficiency, a perturbative argument
confirms that unique attractive states depend in a continuous fashion
on the model parameters \cite{ticozzi-QDS}.

Further work is needed in order to establish 
feedback stabilization results which include finite bandwidth and
detection efficiency, as well as simultaneous monitoring of multiple
observables. In addition, the analysis of Markovian stabilization
problems in the presence of control resources different and/or more
constrained than assumed here appears especially well worth pursuing.
For instance, as illustrated in the last examples, one may want to
limit possible operations to local or collective observables of
multipartite systems in various settings of physical relevance.  The
latter may include opto-mechanical systems, for which feedback control
strategies based on homodyne detection have been considered before
\cite{vitali-optomechanical}, or non-equilibrium many-body systems,
for which preparation of a class of entangled states using
\vv{quasi-local} Markovian dissipation has recently being investigated
in a physically motivated setting \cite{Kraus-entanglement}.
Additional investigation is also required to establish the full
potential of Hamiltonian control and Markovian feedback in
synthesizing not only invariant and attractive, but also {\em
noiseless} structures \cite{ticozzi-QDS}. This may point to yet new
venues for producing protected realizations of quantum information in
physical systems described by quantum Markovian semigroups.

\bibliography{bibliography} 
\bibliographystyle{unsrt}



\end{document}